\documentclass{aa}
\usepackage{graphicx}
\usepackage{natbib}

\begin{document}

\title{The Solar Radius in the EUV during the Cycle XXIII}

\author{C.G. Gim\'enez de Castro \inst{1} 
\and A.C. Varela Saraiva \inst{1,2} 
\and 	J.E.R. Costa \inst{2}
\and 	C.L. Selhorst  \inst{1,2}
}

\institute{Centro de R\'adio Astronomia e Astrof\'{\i}sica Mackenzie, R. da
      Consola\c{c}\~ao 896,  01302-907, S\~ao Paulo, SP, Brazil.
\and  Instituto Nacional de Pesquisas Espaciais, S\~ao Jos\'e dos Campos,
      Brazil.
}

\date{Received 20 June 2007 /  Accepted 29 September 2007}

\titlerunning{UV Solar Radius in Cycle XXIII}

\abstract{} {To determine the solar transition region and coronal radius at
EUV wavelengths and its time evolution during Solar Cycle XXIII.}  {We use
daily 30.4 and 17.1 nm images obtained by the Extreme Ultraviolet Imager
(EIT) aboard the SoHO satellite and derive the solar radius by fitting a
circle to the limb brightness ring.}  {The weighted mean of the temporal
series gives ($967''.56 \pm 0''.04$) and ($969''.54 \pm 0''.02$) at 30.4
and 17.1 nm respectively.  No significant correlation was found with the
solar cycle at any of the two wavelengths.}  {Since the temperature
formation of the 30.4 nm line is between $(60 - 80) \ 10^3$ K (Transition
Region), the obtained result is bigger than that derived from present
optical atmospheric models.  On the contrary this height is
compatible with radio models.} 

\maketitle

\section{Introduction}
The solar atmospheric levels at which most of the different wave-band
brightness are produced have been analyzed by many authors to understand
how the energy flow may affect the local equilibrium. Although there is a
global hydrostatic equilibrium produced by an almost stable photospheric
level, the total solar irradiance (the {\em solar constant}) varies
\cite[e.g.][]{Stix:2002}. The constancy or variation of the
photospheric solar diameter has been of a much debate subject.  The coronal
heating, however, is an indication that some energy is being deposited
above the photosphere modifying the equilibrium.  The chromosphere and
transition region height variations have been verified by solar radius
measurements at radio wavelengths with clear correlation with the solar
constant changes \citep{Costaetal:1999}. \\

In recent years \cite{Emilioetal:2000} and \cite{Kuhnetal:2004} have shown
that the photospheric radius variation, if it exist, must be smaller than
15 mas during a whole solar cycle, therefore it can be taken as constant.
Analysis in other frequencies of the electromagnetic spectrum have been
carried out mainly in the radio domain.  \cite{Selhorstetal:2004} have
reviewed previous results and determined the solar radius variation at 17
GHz using maps from the Nobeyama Radioheliograph (NoRH) with a space
resolution of $5''$. They found a good correlation between the mean radius
and the sunspot number but an anti-correlation between the polar radius and
the sunspot number. \\

At UV frequencies there are very few previous works, with short temporal
series analysis. In one of these very few works, \cite{Zhangetal:1998} used
EIT full Sun images and concluded that: {\em i)} the solar disk is prolate
\cite[the same conclusion was obtained by][]{Auchereetal:1998} and {\em
ii)} the chromospheric 30.4 nm limb is significantly higher than the
limb at 17.1, 19.5 and 28.4 nm wavelengths, arguing that spicules
can contribute to the increase of the chromospheric / Transition Region
(TR) height. \\

The determination of the solar atmosphere layer's height is relevant to the
formulation of atmospheric models.  Many of these models are based on
optical spectral analysis. The density and temperature dependence on height
are derived by fitting a predicted spectrum to the observed one,
e.g. \cite{VAL73,VAL76,VAL81,FAL90,FAL91,FAL93,FAL02}. All of these models
obtain a TR height between 1700 and 2300 km above the photosphere.  On the
contrary, some H$\alpha$ observations show a TR at around 5000 km above
Photosphere \cite[e.g.][]{Zirin:1996}.  At microwaves, the solar radius
has even bigger values $\sim\ 10^4$ km \cite[e.g.][]{Selhorstetal:2004}.  
\citet{Zirin:1996} argued that the hydrostatic
assumption of the atmospheric models is responsible for the disagreement
between predictions and observations.  On the other hand,
\citet{Selhorstetal:2005} claim that spicules are sufficient to increase
the height of the solar limb when a model with a brightness temperature
$T_b \sim 11 \ 000 $ K at 17 GHz and plasma densities few times $10^{10}$
cm$^{-3}$ are used.  \\

In this work we report the determination of the chromospheric/TR and
coronal solar radius and their variation during the solar cycle XXIII using
full Sun images of the Extreme Ultraviolet Imager Telescope (EIT) aboard
the Solar Heliospheric Observatory (SoHO) obtained with the passband
filters centered at 30.4 nm and 17.1 nm.  Our determination is based on the
limb bright ring seen in both wavelengths.  Since the 30.4 nm He line forms
at around few times $10^4$ K we compare our observations with the microwave
determinations of \citet{Selhorstetal:2004}.  We present the method and
results in Sect. \ref{sec:method}, while in Sect. \ref{sec:comp} we analyze
the obtained time series and discuss our results.

\section{The determination of the limb bright ring}
\label{sec:method}

The EIT is one of the 12 instruments aboard the SoHO satellite
\citep{Mosesetal:1997}. It obtains images in four wavelengths by means of
passband filters centered on the coronal lines $\lambda=$~17.1 nm (Fe
{\sc ix,x}, $T\sim\ 10^6$, K), $\lambda=$19.5 nm (Fe {\sc
xii}, $T\sim\ 1.4 \ 10^6$ K) and $\lambda=$~28.4 nm (Fe {\sc xv},
$T\sim\ 2.1 \ 10^6$ K).  The fourth wavelength range corresponds to a
chromospheric / TR line, $\lambda=$~30.4 nm (He~{\sc ii} Ly~$\alpha, 
\ \ T\sim\ 6 - 8 \ 10^4$ K).\\

We choose images centered at the wavelengths $\lambda=$~17.1 nm and
$\lambda=$~30.4 nm.  The first line corresponds to the blending of Fe {\sc
x}, $\lambda=$~17.106 nm and Fe {\sc ix}, $\lambda=$~17.056 nm
\citep{CowanPeacock:1965}. Both lines are optically thin, therefore they
are most intense where the column depth is bigger
\citep[e.g.][]{Withbroe:1970}. The maximum of the Fe~{\sc ix,x} emission is
a pure coronal feature which can be used to determine a solar coronal
radius that can be compared with radio continuum observations.
Furthermore, since the emission is optically thin, changes in the coronal
energy input may affect this radius and therefore this definition allows us
to investigate the coronal heating during a solar cycle. On the other hand,
the maximum of the radial profile is very sharp (see Figure
\ref{fig:fitting}) making the radius determination very precise. Our
criterion for this line is different to the one used by
\cite{Zhangetal:1998} who determines the occulting limb of the EUV
iron lines in a way which is equivalent to \cite{Auchereetal:1998} limb
definition as the inflection point of the radial profile. We note although,
that this occulting or blocking limb corresponds to a Chromospheric layer 
which is investigated in this paper by means of the helium images.\\

The formation of the He {\sc ii} line is more controversial.  There is an
ambiguity between the predicted and observed intensities and a discussion
about its mechanism \citep[e.g.][]{Jordan:1975,Andrettaetal:2003}. In
general the line is considered optically thick and formed in a thin layer
\citep{PietarilaJudge:2004} resulting also in a limb bright ring. Although
this ring is not so intense and clearly defined as the coronal one, it can
be seen in the EIT images during the whole solar cycle. The line is
contaminated by the coronal $\lambda = 30.3$ nm Si~{\sc xi} line, which may
have a relative intensity of up to 20\% above the disk brightening
\citep{Auchere:2000}. Since Si~{\sc xi} is a coronal line, we expect that
it will increase the radius determination, although there is no easy way to
quantify the effect. We believe that since it is a weak line it will
produce a small shift masked by the rest of the uncertainties.\\

\begin{figure*}
\sidecaption
\centerline{
\resizebox{17cm}{!}{\includegraphics{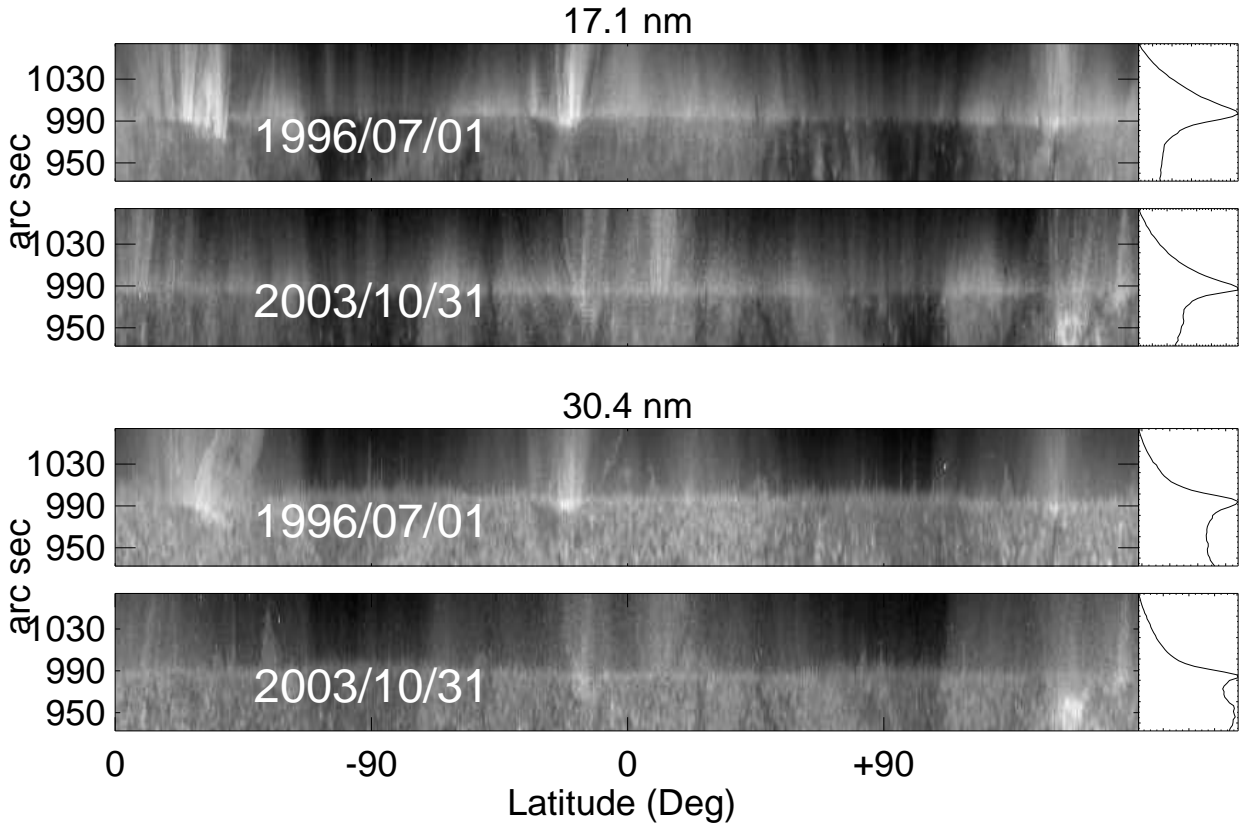}}}
\caption{EUV limb brightening.  Images show the limb during different
solar cycle phases.  Right panels represent the mean limb profile.\label{fig:limb}}
\end{figure*}

We define the solar radius as the semi-distance between the diametrically
opposed maxima of the limb bright points of the rings.  This definition allows us to
determine the radius with an automatic methodology which is explained
below. \\

\subsection{Limb Extraction and Radius Determination}
\label{subsec:extraction}

\begin{figure}
\centerline{
\resizebox{9.5cm}{!}{\includegraphics{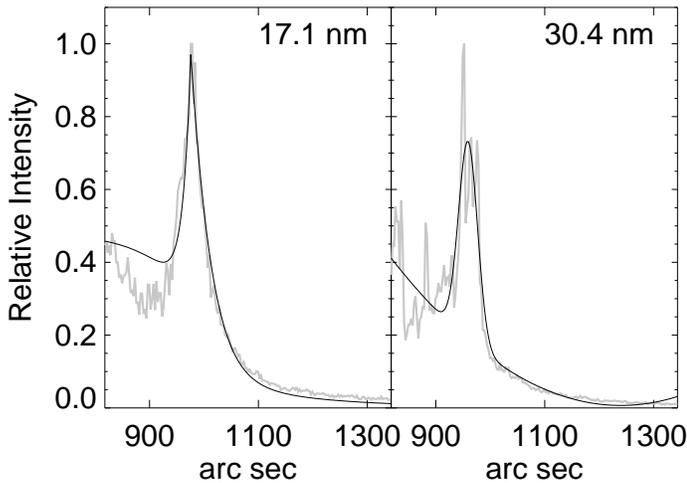}}}
\caption{Examples of limb fitting. Left, a limb profile at 17.1 nm
(gray) and the fitting (black). Right, the same for the 30.4
nm limb profile.\label{fig:fitting}}
\end{figure}

Every image was corrected using standard procedures included in the Solar
Software package for the EIT instrument.  We use the \cite{Canny:1986}
algorithm to get a first approach to the limb ring, a circular fitting to
this ring gives us a disk center determination better than the one found in
the image header. Once the center is well determined, radial profiles are
extracted every half a degree in polar angle. The limb brightening of each
profile is fitted to a special function $f(r)$ that depends on the
wavelength band:
\begin{description}
\item[17.1 nm : ]  $f(r) = a_\circ \ \mathrm{atan}((a_1-r) \ a_2) + a_3 + a_4
\ r^{a_5} \exp(-|r-r_m|\ a_6)$ \quad .
\item[30.4 nm : ]  $f(r) = a_\circ \ \exp{(-(r-r_m)^2/2a_1^2)}+a_2 + a_3 \ r +
a_4 \ r^2 $ \quad . 
\end{description}
Fe~{\sc ix,x} limb brightening is better represented by a cusp function
while the arctangent is used to represent the transition between the
scattered coronal light and the quiet sun.  He~{\sc ii} limb profiles do
not have a clear peak and thus are not well represented by a cusp function,
therefore we use a Gaussian to determine the mean brightest position of the
limb.  In this case a second degree polynomial represents the transition
from the scattered light and the quiet sun. In both functions, $a_i$ and
$r_m$ are the fitting parameters to be determined, the latter is the
position of the limb maximum. The fitting is carried out by means of the
amoeba algorithm \citep{Pressetal:2002}, setting the optical radius as the
lower limit of the fitting interval (see Figure \ref{fig:fitting} for
examples of the fittings). We iteratively fit a circle to the set of
$(x,y)$ ring positions discarding at every step those that are
outside the band $(R - 1.8 \ \sigma, R + 1.8 \ \sigma)$, with $R$ the
circle radius and $\sigma$ the standard deviation of the fitting. The
iteration goes until it converges with all pairs inside the fitting
band. To convert from the CCD pixel units to arc seconds ($''$), we use the
plate scale determined during a 1999 Mercury transit by
\cite{AuchereArtzner:2004}, namely $2.627\pm 0.001 \ \mathrm{arc \sec \
pixel}^{-1}$. This figure was later confirmed with a new transit of Mercury
in 2003 (Artzner, personal communication).  The resulting size is corrected
to 1 AU using the satellite distance found in the image header.\\

\subsection{Results}
\label{subsec:results}

\begin{figure}
\centerline{
\resizebox{9.5cm}{!}{\includegraphics{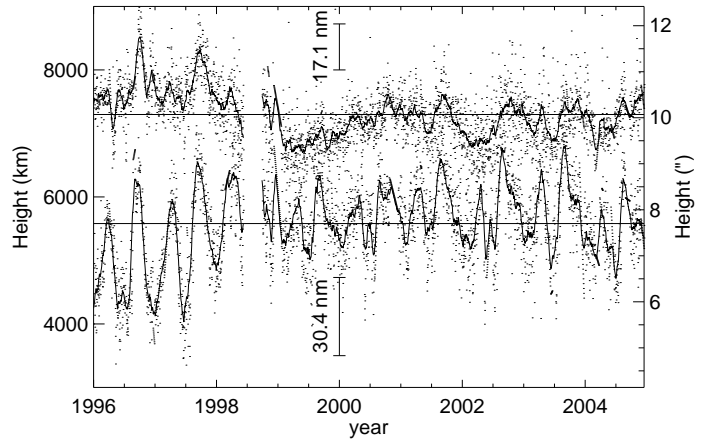}}}
\caption{Time evolution of the solar radii.  Points represent individual
measurements.  Solids curves are 28 days running means at 17.1 nm (upper)
and 30.4 nm (lower). Straight thick lines are the weighted mean values for
the period 1999 - 2005. The vertical bars represent the mean uncertainty
for one measurement. \label{fig:time-profile}}
\end{figure}

Figure \ref{fig:time-profile} shows the time evolution of the solar 17.1
and 30.4 nm radii.  Heights are relative to the photospheric radius at 1
AU, $R_\odot = 959''.68 = 695489.2 \ $ km.  Points represent individual
measurements, continuous curves are 28 days running means and the
horizontal thick lines are the weighted means. There are some gaps in the
data when the instrument stoped its operation. Remarkable oscillations are
observed before the gap at the end of 1998. An incorrect determination of
the satellite distance, or a variation of the telescope focal distance
produced by thermal oscillations can account for an incomplete data
correction.  Nonetheless it is not clear why the effect is so important at
30.4 nm and less significant at 17.1 nm.  Since we do not have certainty on
the origin of the strong oscillations we decide not to use the data for the
analysis before the long observing stop of the instrument in 1998. In
general, the mean uncertainty in radius determination is $1''.7$, almost
half a pixel, at 30.4 nm and $1''$, a third of a pixel, at 17.1 nm. The
bigger uncertainty for the line of He~{\sc ii} reveals its weaker intensity
and its complex nature.\\

The very interesting harmonic variations of the solar limb during the cycle
shown in \ref{fig:time-profile} are still under analyses to be published
somewhere else.

\section{Analysis and Discussion}
\label{sec:comp}

\begin{table}
\begin{tabular}{ccc}
  & \bfseries Radius & \bfseries Height \\
  & $('')$   & km \\
\hline\hline
\bfseries He~{\sc ii}   & $967.56 \ \pm \ 0.04$ & $5710 \ \pm 30 $ \\
\bfseries Fe~{\sc ix,x} & $969.54 \ \pm \ 0.02$ & $7146 \ \pm 15 $ \\
\hline
\end{tabular}
\caption{\label{tbl:results} Solar radius and height over the
  photosphere.}
\end{table}

Table \ref{tbl:results} summarizes our results. First we note that the
weighted means show a clear different height for the chromospheric He~{\sc
ii} and the coronal Fe~{\sc ix,x} lines. Even if we use the mean
uncertainty of one measurement the heights do not overlap.  It is also
clear that the chromosphere height is well above where present atmospheric
models predict and near where \cite{JohannessonZirin:1996} found it using
H$\alpha$ filtergrams and is comparable to \cite{Zhangetal:1998}.\\

\cite{Auchereetal:1998} have shown that the Sun is $1''.5$ and $5''$
prolate at 17.1 and 30.4 nm respectively. \cite{Zhangetal:1998} have also
observed a similar prolatness of the EUV solar limb. Due to the noisy limb
we did not fit an ellipse. We use instead a circle fit much simpler and
faster.  It should be noted that the ellipsoidal form is revealed after the
average of over 400 images in \cite{Auchereetal:1998} work. Therefore, our
results represent an averaged value over the solar limb.  Taking Auchere's
results into account we can correct our determination of the equatorial
radius subtracting $2''$ and the polar radius adding $3''$.  In that way we
get 4130~km equatorial height and 7750~km polar
height. \cite{Zhangetal:1998} find $3100\pm 1200$ km and $6600\pm 1200$ km
for the equatorial and polar heights respectively, in accordance with our
results. On the other hand, we cannot compare our 17.1 nm height with that
obtained in Zhang's work because of our different definition of coronal
radius.\\ \\

\begin{figure}
\centerline{
\resizebox{!}{12cm}{\includegraphics{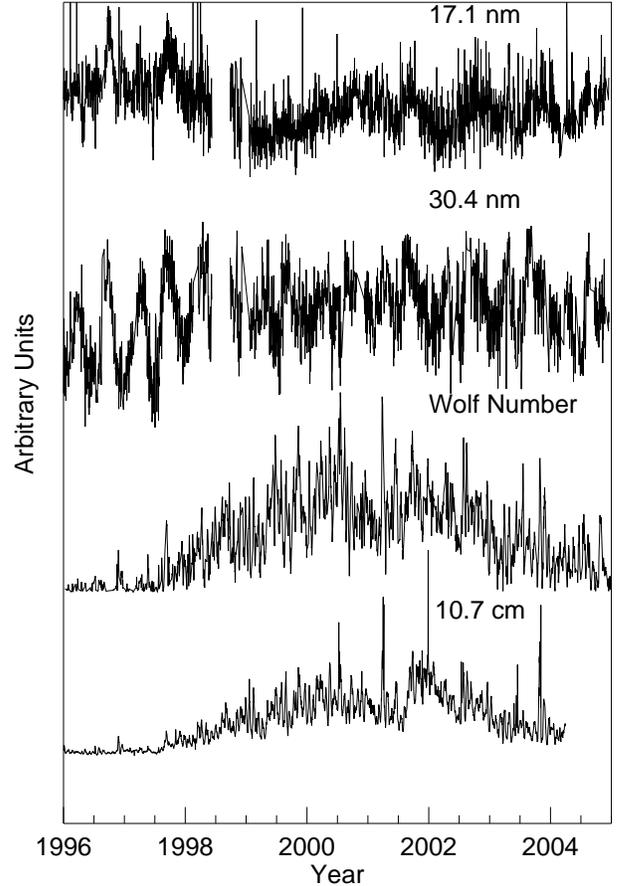}}}
\caption{EUV radius variations compared with solar cycle indices. From top
to bottom: 17.1 nm radius, 30.4 nm radius, Wolf Number and flux density at
10.7 cm.  All values are rescaled between $[-0.5,0.5]$ to facilitate the
comparison. \label{fig:comparaciones}}
\end{figure}

\subsection{Comparison with radio observations and solar indices variations}
\label{subsec:compaaraciones}

The SSC atmospheric model proposed by \cite{Selhorstetal:2005} predicts a
solar radius of $966''.5$ at 17 GHz.  This microwave continuum emission
comes from a solar atmospheric layer at a temperature of around $10^4$ K,
i.e. the chromosphere below the region where the Fe~{\sc ix,x} and He~{\sc
ii} are formed. Thus, the results of this work agree with that model. Not
withstand, \cite{Selhorstetal:2004} have determined a radius of around
$975''$ from 17 GHz NoRH maps. In \cite{Selhorstetal:2005} the authors
claim that the presence of spicules can explain the higher observed radius.
Since typical spicules temperature is between $0.5 \ 10^4$ and $2.5 \ 10^4$
K \citep[see][and references therein]{Sterling:2000}, their contribution to
the EUV lines can be neglected and there is no contradiction in having the
height of the 30.4 nm chromospheric line forming below the observed 17 GHz
continuum. \\

In Figure \ref{fig:comparaciones} we plot our results together with the
daily Wolf Number and the daily mean flux density at 10.7 cm solar indices.
To ease the comparison we have rescaled the values in the band
$[-0.5,0.5]$. The figure shows no clear indication of a close association
between the radius and the solar indices.  The cross correlation between
the He~{\sc ii} and Fe~{\sc ix,x} radius and the solar indices are shown in
table \ref{tbl:xcorr}. The near zero figures indicate that no correlation
exists, although one is tempted to affirm that there is some anti
correlation, of around -0.4, between the coronal line and the solar
indices. The real nature of such an anti correlation remains to be
confirmed. In any case, it is surprising that the solar activity does not
change the solar EUV radius neither at chromospheric nor at coronal heights
as it is observed, v.g. at 17 GHz \citep{Selhorstetal:2005}. As an
explanation we think that the microwave continuum comes from a wide
atmospheric layer receiving an important contribution from the spicules
while a narrower layer is expected to form the EUV lines.  Therefore it is
possible that the variation in position of the EUV layers are not
detectable with the present methodology and instrumentation. \\

\begin{table}
\begin{tabular}{ccccccc}\hline
  & \bfseries 17 GHz & \bfseries Irr. & \bfseries Wolf N. & \bfseries
  10.7 cm\\[0.5ex]
\hline\hline
{\bfseries He {\sc ii}}      &  +0.14   &  +0.20   &  +0.30   &  -0.12\\
{\bfseries Fe {\sc ix,x}}    &  -0.12   &  -0.38   &  -0.43   &  -0.40\\
\hline
\end{tabular}\\
\caption{\label{tbl:xcorr} Cross correlation between our results of
 the solar radii and solar activity indicators. From left to right:
 solar radius at 17 GHz, Total Solar Irradiance, Wolf Number and flux
 density daily means at 10.7 cm (S-component). }
\end{table}

Although we find no long term variations and the use of the running means
erases changes with durations of the order or lower than 28 days, Figure
\ref{fig:time-profile} shows fluctuations with hundred days temporal scale,
more sensitive for the He~{\sc ii} than for the Fe~{\sc ix,x} images. 
This variation deserves a careful examination and a future study will have 
to show if it is just an instrumental effect or a real solar phenomenon.\\

\begin{acknowledgements}
ACS acknowledges support from the Funda\c{c}\~ao de Amparo \`a Pesquisa do
estado de S\~ao Paulo (FAPESP) under grant number 03/03500-2.  The present
project received partial support from Fundo Mackenzie de Pesquisa
(Mackpesquisa).  The authors are in debt with G. Artzner and M. Emilio
because their fruitful discussions and informations.
\end{acknowledgements}

\bibliographystyle{aa}

\end{document}